\documentclass[11pt]{revtex4}    
\usepackage{amssymb,epsf}
\usepackage{latexsym}
\begin{document}

\title{Charged rotating dilaton black strings in AdS spaces}
\author{Ahmad Sheykhi \footnote{sheykhi@mail.uk.ac.ir}}
\address{Department of Physics, Shahid Bahonar University, P.O. Box 76175, Kerman, Iran\\
         Research Institute for Astronomy and Astrophysics of Maragha (RIAAM), Maragha, Iran}
\begin{abstract}
We derive a class of charged rotating dilaton black string
solutions in the background of anti-de-Sitter spaces with an
appropriate combination of three Liouville-type dilaton
potentials. We also present the suitable counterterm which removes
the divergences of the action in the presence of dilaton
potential. The solutions are analyzed and their thermodynamics is
discussed by using the counterterm method.
\end{abstract}
\maketitle

There has been considerable attention in the past years in dilaton
gravity. It is of great importance to investigate the effect of
the dilaton field on the properties of the solutions. It was found
that the dilaton field changes the causal structure of the black
hole and leads to the curvature singularities at finite radii. In
the absence of dilaton potential, exact solutions of charged
dilaton black holes have been constructed by many authors
\cite{CDB1,CDB2}. These black holes are all asymptotically flat.
The presence of Liouville-type dilaton potential, which is
regarded as the generalization of the cosmological constant,
changes the asymptotic behavior of the solutions to be neither
asymtotically flat nor (anti)-de Sitter [(A)dS]. Indeed, it has
been shown that with the exception of a pure cosmological
constant, no dilaton de-Sitter or anti-de-Sitter black hole
solution exists with the presence of only one Liouville-type
dilaton potential \cite{MW}. In the presence of one or two
Liouville-type potential, black hole spacetimes which are neither
asymptotically flat nor (A)dS have been explored by many authors
(see e.g. \cite{CHM,Cai,Clem,Shey0}). Although these kind of
solutions may shed some light on the possible extensions of
AdS/CFT correspondence, they are physically less interesting due
to their unusual asymptotic behavior.

On the other side, the construction and analysis of black hole
solutions in (A)dS  space is a subject of much recent interest.
This is primarily due to their relevance for the AdS/CFT
correspondence. It was argued that the thermodynamics of black
holes in AdS spaces can be identified with that of a certain dual
CFT in the high temperature limit \cite{Witt}. Having the AdS/CFT
correspondence idea at hand, one can gain some insights into
thermodynamic properties and phase structures of strong 't Hooft
coupling CFTs by studying thermodynamics of AdS black holes.
Recently, the dilaton potential leading to (anti)-de Sitter-like
solutions of dilaton gravity has been found \cite{Gao1} (see also
\cite{Gao2}). It was shown that the cosmological constant is
coupled to the dilaton in a very nontrivial way. With the
combination of three Liouville-type dilaton potentials, a class of
static dilaton black hole solutions in (A)dS spaces has been
obtained by using a coordinates transformation which recast the
solution in the schwarzschild coordinates system \cite{Gao1}. The
purpose of the present Letter is to construct a class of charged
rotating dilaton black string solutions in the background of AdS
spacetime. We will also present the suitable counterterm which
removes the divergences of the action. Finally, we analyze the
solutions and calculate their conserved and thermodynamic
quantities by using the counterterm method inspired by AdS/CFT
correspondence.

\begin{figure}[tbp]
\epsfxsize=7cm \centerline{\epsffile{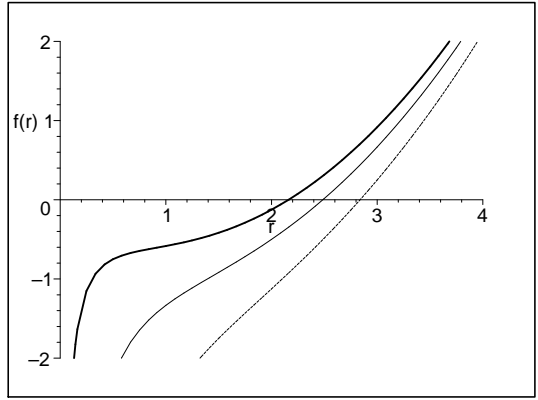}} \caption{The
function $f(r)$ versus $r$ for $\protect\alpha=1$, $b=2$ and
$\Lambda=-1$. $q=0.5$ (bold line), $q=1$ (continuous line) and
$q=1.5$ (dashed line).} \label{figure1}
\end{figure}

\begin{figure}[tbp]
\epsfxsize=7cm \centerline{\epsffile{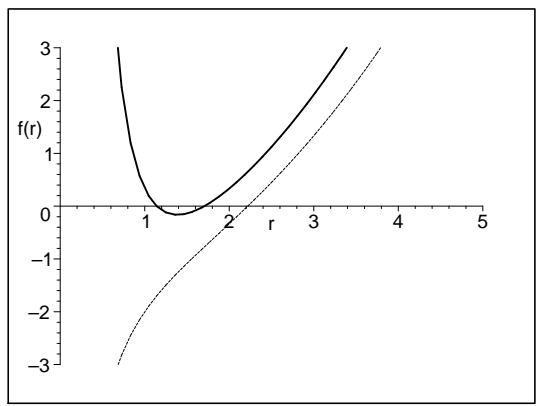}} \caption{The
function $f(r)$ versus $r$ for $b=1$, $q=2$ and $\Lambda=-1$.
$\protect\alpha=0$ (bold line), $\protect\alpha=1$ (dashed line).}
\label{figure2}
\end{figure}

Our starting point is the four-dimensional
Einstein-Maxwell-dilaton action
\begin{eqnarray}
I_{G} &=&-\frac{1}{16\pi }\int_{\mathcal{M}}d^{4}x\sqrt{-g}\left(
R\text{ }-2\partial_{\mu}\Phi \partial^{\mu}\Phi-V(\Phi
)-e^{-2\alpha \Phi
}F_{\mu \nu }F^{\mu \nu }\right)   \nonumber \\
&&-\frac{1}{8\pi }\int_{\partial \mathcal{M}}d^{3}x\sqrt{-\gamma
}\Theta (\gamma ),  \label{Act}
\end{eqnarray}
where ${R}$ is the Ricci scalar curvature, $\Phi $ is the dilaton
field, and $V(\Phi )$ is a potential for $\Phi $. $\alpha $ is a
constant determining the strength of coupling of the scalar and
electromagnetic field, $F_{\mu \nu }=\partial _{\mu }A_{\nu
}-\partial _{\nu }A_{\mu }$ is the electromagnetic  field tensor
and $A_{\mu }$ is the electromagnetic potential. The last term in
Eq. (\ref{Act}) is the Gibbons-Hawking surface term. It is
required for the variational principle to be well-defined. The
factor $\Theta$ represents the trace of the extrinsic curvature
for the boundary ${\partial \mathcal{M}}$ and $\gamma$ is the
induced metric on the boundary. While $\alpha=0$ corresponds to
the usual Einstein-Maxwell-scalar theory, $\alpha=1$ indicates the
dilaton-electromagnetic coupling that appears in the low energy
string action in Einstein's frame. In this Letter, we examine
action (\ref{Act}) with three Liouville-type dilaton potentials
\cite{Gao1}
\begin{equation}
V(\Phi )=\frac{2\Lambda}{3(\alpha^2+1)^{2}}\left[ {\alpha}^{2}
\left( 3\,{\alpha}^{2}-1 \right) {e ^{-2\Phi/\alpha}}+ \left(
3-{\alpha}^{2} \right) {e^{2\,
\alpha\,\Phi}}+8\,{\alpha}^{2}{e^{\Phi(\alpha-1/\alpha)}} \right]
, \label{v1}
\end{equation}
where $\Lambda $ is the cosmological constant. It is clear the
cosmological constant is coupled to the dilaton in a very
nontrivial way. This type of dilaton potential can be obtained
when a higher dimensional theory is compactified to four
dimensions, including various supergravity models \cite{Gid}. The
equations of motion can be obtained by varying the action
(\ref{Act}) with respect to the gravitational field $g_{\mu \nu
}$, the dilaton field $\Phi $, and the gauge field $A_{\mu }$
which yields the following field equations
\begin{equation}
{R}_{\mu \nu }=2 \partial _{\mu }\Phi \partial _{\nu }\Phi
+\frac{1}{2}g_{\mu \nu }V(\Phi )+2e^{-2\alpha \Phi}\left( F_{\mu
\eta }F_{\nu }^{\text{ }\eta }-\frac{1}{4}g_{\mu \nu }F_{\lambda
\eta }F^{\lambda \eta }\right) ,  \label{FE1}
\end{equation}
\begin{equation}
\nabla ^{2}\Phi =\frac{1}{4}\frac{\partial V}{\partial \Phi
}-\frac{\alpha }{2}e^{-2\alpha \Phi}F_{\lambda \eta }F^{\lambda
\eta }, \label{FE2}
\end{equation}
\begin{equation}
\partial _{\mu }\left( \sqrt{-g}e^{-2\alpha \Phi}F^{\mu \nu
}\right) =0.  \label{FE3}
\end{equation}
Our aim here is to construct charged rotating black string
solutions of the field equations (\ref{FE1})-(\ref{FE3}) and
investigate their properties. The metric of four-dimensional
rotating solution with cylindrical or toroidal horizons can be
written as \cite{Lem}
\begin{eqnarray}\label{metric}
ds^{2} &=&-f(r)\left( \Xi dt-a d\phi \right) ^{2}+%
r^{2}R^{2}(r)\left(\frac{a}{l^2} dt-\Xi d\phi \right) ^{2}+\frac{dr^{2}}{f(r)}+\frac{r^{2}}{l^{2}}R^{2}(r)dz^{2},  \nonumber \\
\Xi ^{2} &=&1+\frac{a^{2}}{l^{2}},  \label{Met3}
\end{eqnarray}
where $a$ is the rotation parameter. The functions $f(r)$ and
$R(r)$ should be determined and $l$ has the dimension of length
which is related to the constant $\Lambda $ by the relation
$l^{2}=-3/\Lambda $. The two dimensional space, $t$=constant and
$r$ =constant, can be (i) the flat torus model $T^2$ with topology
$S^1 \times S^1$, and $0\leq \phi<2\pi$, $0\leq z<2\pi l$, (ii)
the standard cylindrical model with topology $R\times S^1$, and
$0\leq \phi<2\pi$, $-\infty< z<\infty$,  and (iii) the infinite
plane $R^2$ with $-\infty< \phi<\infty$ and $-\infty< z<\infty$.
We will focus upon (i) and (ii). The Maxwell equation (\ref{FE3})
can be integrated immediately to give
\begin{eqnarray}
F_{tr} &=&\frac{q\Xi e^{2\alpha \Phi}}{r^2R^2(r)}, \nonumber
\label{Ftr} \\
F_{\phi r} &=&-\frac{a}{\Xi }F_{tr},
\end{eqnarray}
where $q$, an integration constant, is related to the electric
charge of black string. Inserting the Maxwell fields (\ref{Ftr})
and the metric (\ref{Met3}) in the field equations (\ref{FE1}) and
(\ref{FE2}), we can write these equation for $a=0$ as
\begin{eqnarray}
&&2r^4 R ^3 R' f'+ r^4 R ^4 f'' +2r^3 R ^4 f'+r^4 R ^4 V(
\Phi)-2q^2 e^{2\alpha\Phi}=0, \label{Eqtt} \\&& 2r^3 R ^4 f'+ r^4
R ^4 f'' +8r^3 R ^3 f R'+4r^4 R ^3 f R''+2r^4
R ^3 R' f' \nonumber \\
&&+4 r^4 R ^4 f \Phi'^2+ r^4 R ^4 V \left(
\Phi \right) -2q^2 e^{2\alpha\Phi}=0, \label{Eqrr} \\
&& 8r^3 R ^3 f R'+ 2r^3 R^4
f'+ 2r^4 R ^3 f' R'+2r^2 R ^4 f+2r^4 R ^2 f R'^2 \nonumber \\
&&+2r^4 R^3 f R''+ r^4 R ^4 V \left( \Phi \right) +2q^2
e^{2\alpha\Phi}=0, \label{Eqpp}\\
&& r^4 R ^4 {\Phi'}{f'}+r^4 R ^4
{\Phi''} {f}+2r^3 R ^4 \Phi' f+2 r^4 R ^3 R' {\Phi'}{f}- r^4 R ^4
\frac {\partial{V}}{4\partial{\Phi}}-\alpha q^2
e^{2\alpha\Phi}=0,\label{Eq3}
\end{eqnarray}
where the ``prime" denotes differentiation with respect to $r$.
Subtracting Eq. (\ref{Eqrr}) from Eq. (\ref{Eqtt}) gives
\begin{eqnarray}\label{Eqtt-Eqrr}
2R' +r R'' + r R \Phi'^2=0.
\end{eqnarray}
Then we make the ansatz
\begin{equation}\label{Rphi}
 R(r)=e^{\alpha \Phi}.
\end{equation}
Substituting this ansatz in Eq. (\ref{Eqtt-Eqrr}), it reduces to
\begin{eqnarray}\label{Eqphi}
r \alpha \Phi''+2\alpha \Phi'+r(1+{\alpha}^{2}) \Phi'^2=0,
\end{eqnarray}
which has a solution of the form
\begin{equation}
\Phi (r)=\frac{\alpha }{\alpha ^{2}+1}\ln (1-\frac{b}{r}),
\label{phi}
\end{equation}
where $b$ is an integration constants. Inserting (\ref{phi}), the
ansatz (\ref{Rphi}), and the dilaton potential (\ref{v1}) into the
field equations (\ref{Eqtt})-(\ref{Eq3}), one can show that these
equations have the following solution
\begin{equation}\label{f(r)}
f(r)=-\frac{c}{r} \left( 1-{\frac {b}{r}} \right) ^{{\frac
{1-{\alpha}^{2}}{1+{\alpha}^ {2}}}}-\frac{\Lambda}{3}\,{r}^{2}
\left( 1-{\frac {b}{r}}
 \right) ^{{\frac {2{\alpha}^{2}}{{\alpha}^{2}+1}}},
\end{equation}
where $c$ is an integration constant. The above solutions will
fully satisfy the system of equations (\ref{Eqtt})-(\ref{Eq3})
provided we have $q^2(1+\alpha^2)=bc$. One can also check that
these solutions satisfy equations (\ref{FE1})-(\ref{FE3}) in the
rotating case where $a\neq0$. It is apparent that this spacetime
is asymptotically (anti)-de-Sitter. The Kretschmann scalar $R_{\mu
\nu \lambda \kappa }R^{\mu \nu \lambda \kappa }$  and the Ricci
scalar $R$ diverge at $r=0$ and therefore there is an essential
singularity located at $r=0$. The explicit form of the Kretschmann
scalar is complicated and we do not present it here, however, the
Ricci scalar curvature of the metric has a simpler form and can be
written as
\begin{equation}\label{Ricci}
R=-\frac{2\,{q}^{2} {\alpha}^{2}b \left( 1-{\frac {b}{r}} \right)
^{{\frac { 1- \alpha^2}{{1+\alpha}^{2}}}}}{
 \left( r-b \right) ^{2} \left( {\alpha}^{2}+1 \right) {r}^{3}}+
\frac{2\Lambda\, \left( 1-{\frac {b}{r}} \right) ^{{\frac
{2{\alpha}^{2}}{{\alpha }^{2}+1}}} \left( 2\left( r-b
 \right) ^{2}+4\,{\alpha}^{2}{r}(r-b)+2\,{\alpha}^{4}{r}^{2}-{\alpha}^{2}{b}^{2}
 \right) } {\left( {\alpha}^{2}+1 \right) ^{2} \left( r-b
 \right) ^{2}}.
\end{equation}
One can easily check that for the arbitrary values of the dilaton
coupling constant,  $R\rightarrow 4\Lambda$  as $r\rightarrow
\infty$. We also find out that for arbitrary $\alpha$ the
Kretschmann scalar approach $ 8\Lambda^2/3$ as $r\rightarrow
\infty$. This confirm our above discussion that our solution is
asymptotically anti-de-Sitter. In the absence of a nontrivial
dilaton ($\alpha=0$), the solution reduces to the asymptotically
(anti)-de-Sitter charged rotating black string \cite{Lem}.
However, for $\alpha\neq0$ the solution is qualitatively
different. As one can see from Eq. (\ref{Ricci}), the surface $r =
b$ is a curvature singularity except for the case $\alpha = 0$
when it is a nonsingular inner horizon. This is consistent with
the idea that the inner horizon is unstable in the
Einstein-Maxwell theory. Therefore, our solutions describe black
strings in the case $b< r_{+}$, where $r_{+}$ is the outer horizon
of the black string (the root of Eq. $ f(r)=0$). The above
discussions will become more clear  if we look on the figures
\ref{figure1} and \ref{figure2}, where we have plotted the
function $f(r)$ versus $r$ for different values of the dilaton
coupling $\alpha$ and the charge parameter $q$. In the case $r=
b$, it is clear from (\ref{metric}) and (\ref{Rphi}) that for
$\alpha \neq 0$  the area of the event horizon goes to zero (since
$R(r)\rightarrow 0$ in this case).

In summary, compared to the charged black holes/strings in (A)dS
universe, the dilaton version has some remarkable properties. In
the first place, there may be three horizons in the
Reissner–Nordström–de Sitter spacetime, i.e., black hole event
horizon, black hole Cauchy horizon and cosmic event horizon.
However, the charged dilaton black holes/strings in (A)dS spaces
has at most two horizons. Here the inner Cauchy horizon
disappears. This is due to the fact that the inner horizon is
unstable, as pointed by Garfinkle \cite{CDB2}.

Next, we calculate the conserved quantities of the solutions. For
asymptotically (anti)-de-Sitter solutions, the way that one can
calculate these quantities and obtain finite values for them is
through the use of the counterterm method inspired by (A)dS/CFT
correspondence \cite{Mal}. In this paper we deal with the
spacetimes with zero curvature boundary, $R_{abcd}(\gamma )=0$,
and therefore the counterterm for the stress energy tensor should
be proportional to $\gamma ^{ab}$. We find the suitable
counterterm which removes the divergences in the form
\begin{equation}\label{cont}
 I_{ct}=-\frac{1}{8\pi }\int_{\partial \mathcal{M}}d^{3}x\sqrt{-\gamma
}\left(-\frac{1}{l}+\frac{\sqrt{-6V(\Phi)}}{2} \right).
\end{equation}
One may note that in the absence of a dilaton field where we have
$V(\Phi)=2\Lambda=-6/l^2$, the above counterterm has the same form
as in the case of asymptotically (A)dS solutions with
zero-curvature boundary. Having the total finite action $I =
I_{G}+I_{\mathrm{ct}}$ at hand, one can use the quasilocal
definition  to construct a divergence free stress-energy tensor
\cite{BY}. Thus the finite stress-energy tensor in four
dimensional Einstein-dilaton gravity with three Liouville-type
dilaton potentials (\ref{v1}) can be written as
\begin{equation}
T^{ab}=\frac{1}{8\pi }\left[ \Theta ^{ab}-\Theta \gamma
^{ab}+\left(-\frac{1}{l}+\frac{\sqrt{-6V(\Phi)}}{2} \right)\gamma
^{ab}\right] ,  \label{Stres}
\end{equation}
The first two terms in Eq. (\ref{Stres}) are the variation of the
action (\ref{Act}) with respect to $\gamma _{ab}$, and the last
two terms are the variation of the boundary counterterm
(\ref{cont}) with respect to $\gamma _{ab}$. To compute the
conserved charges of the spacetime, one should choose a spacelike
surface $ \mathcal{B}$ in $\partial \mathcal{M}$ with metric
$\sigma _{ij}$, and write the boundary metric in ADM
(Arnowitt-Deser-Misner) form:
\[
\gamma _{ab}dx^{a}dx^{a}=-N^{2}dt^{2}+\sigma _{ij}\left( d\varphi
^{i}+V^{i}dt\right) \left( d\varphi ^{j}+V^{j}dt\right) ,
\]
where the coordinates $\varphi ^{i}$ are the angular variables
parameterizing the hypersurface of constant $r$ around the origin, and $N$
and $V^{i}$ are the lapse and shift functions respectively. When there is a
Killing vector field $\mathcal{\xi }$ on the boundary, then the quasilocal
conserved quantities associated with the stress tensors of Eq. (\ref{Stres})
can be written as
\begin{equation}
Q(\mathcal{\xi )}=\int_{\mathcal{B}}d^{2}x \sqrt{\sigma }T_{ab}n^{a}%
\mathcal{\xi }^{b},  \label{charge}
\end{equation}
where $\sigma $ is the determinant of the metric $\sigma _{ij}$, $\mathcal{%
\xi }$ and $n^{a}$ are, respectively, the Killing vector field and
the unit normal vector on the boundary $\mathcal{B}$. For
boundaries with timelike ($\xi =\partial /\partial t$) and
rotational ($\varsigma =\partial /\partial \varphi $) Killing
vector fields, one obtains the quasilocal mass and angular
momentum
\begin{eqnarray}
M &=&\int_{\mathcal{B}}d^{2}x \sqrt{\sigma }T_{ab}n^{a}\xi ^{b},
\label{Mastot} \\
J &=&\int_{\mathcal{B}}d^{2}x \sqrt{\sigma }T_{ab}n^{a}\varsigma
^{b}.  \label{Angtot}
\end{eqnarray}
These quantities are, respectively, the conserved mass and angular
momenta of the system enclosed by the boundary $\mathcal{B}$. Note
that they will both depend on the location of the boundary
$\mathcal{B}$ in the spacetime,
although each is independent of the particular choice of foliation $\mathcal{%
B}$ within the surface $\partial \mathcal{M}$. The mass and
angular momentum per unit length of the string when the boundary $
\mathcal{B}$ goes to infinity can be calculated through the use of
Eqs. (\ref{Mastot}) and (\ref{Angtot}). We find
\begin{eqnarray}\label{M}
{M}&=&\frac{\alpha^2(\alpha^2-1)b^{3}}{24\pi l^3(\alpha
^{2}+1)^3}+ \frac{(3\Xi ^{2}-1)c}{16\pi l},
\end{eqnarray}
\begin{eqnarray}
{J}&=&\frac{3\Xi c\sqrt{\Xi^2-1}}{16\pi}.  \label{J}
\end{eqnarray}
For $a=0$ ($\Xi =1$), the angular momentum per unit volume
vanishes, and therefore $a$ is the rotational parameters of the
spacetime. The entropy of the  dilaton black string still obeys
the so called area law of the entropy which states that the
entropy of the black hole is a quarter of the event horizon area
\cite{Beck}. This near universal law applies to almost all kinds
of black holes, including dilaton black holes, in Einstein gravity
\cite{hunt}. It is a matter of calculation to show that the
entropy per unit length of the string is
\begin{equation}\label{ent}
{S}=\frac{r_{+}^2
\Xi(1-\frac{b}{r_{+}})^{\frac{2\alpha^2}{\alpha^2+1}}}{4l}.
\end{equation}
By analytic continuation of the metric we can obtain the
temperature and angular velocity of the horizon. The analytical
continuation of the Lorentzian metric by $t\rightarrow i\tau$ and
$a\rightarrow ia$ yields the Euclidean section, whose regularity
at $r = r_+$ requires that we should identify $\tau\sim
\tau+\beta_+$ and $\phi\sim\phi+i\beta _+\Omega_+$ where $\beta_+$
and $\Omega_+$ are the inverse Hawking temperature and the angular
velocity of the horizon. We find
\begin{eqnarray}\label{Tem}
T_{+}&=&\frac{f^{\text{ }^{\prime }}(r_+)}{4\pi \Xi} =\frac{\left(
1-{\frac {b}{r_+}} \right) ^{{\frac {{\alpha}^{2}-1}{{
\alpha}^{2}+1}}}}{2\pi r_+^2\Xi({\alpha}^{2}+1)} \left(
-\frac{\Lambda }{3}r_+^2\left( {\alpha}^{2}r_{+}+r_{+}-b
 \right)+\frac{c\left( {\alpha}^{2}r_{+}+r_{+}-2\,b
\right)}{2r_+} \left(1-{\frac {b}{r_+}} \right) ^{{\frac {1-3
\,{\alpha}^{2}}{{\alpha}^{2}+1}}} \right) \nonumber \\
\Omega_+&=&\frac{a}{\Xi l^2}.
\end{eqnarray}
The next quantity we are going to calculate is the electric charge
of the string. To determine the electric field we should consider
the projections of the electromagnetic field tensor on special
hypersurface. The normal vectors to such hypersurface are
\begin{equation}
u^{0}=\frac{1}{N},\text{ \ }u^{r}=0,\text{ \
}u^{i}=-\frac{V^{i}}{N},
\end{equation}
where $N$ and $V^{i}$ are the lapse function and shift vector.
Then the electric field is $E^{\mu }=g^{\mu \rho }e^{-2\alpha
\phi}F_{\rho \nu }u^{\nu }$, and the electric charge per unit
length of the string can be found by calculating the flux of the
electric field at infinity, yielding
\begin{equation}
{Q}=\frac{\Xi q}{4\pi l}.  \label{Q}
\end{equation}
The electric potential $U$, measured at infinity with respect to
the horizon, is defined by \cite{Cvetic}
\begin{equation}\label{U}
U=A_{\mu }\chi ^{\mu }\left| _{r\rightarrow \infty }-A_{\mu }\chi
^{\mu }\right| _{r=r_{+}},
\end{equation}
where $\chi =\partial _{t}+\Omega \partial _{\phi}, $ is the null
generator of the event horizon. One can easily show that the
vector potential $A_{\mu }$ corresponding to the electromagnetic
tensor (\ref{Ftr}) can be written as
\begin{equation}
A_{\mu }=-\frac{q}{r}\left( \Xi \delta _{\mu }^{t}-a\delta _{\mu
}^{\phi}\right). \label{Pot}
\end{equation}
Substituting (\ref{Pot}) in (\ref{U}) we obtain the electric
potential as
\begin{equation}
U=\frac{q}{\Xi r_{+}}. \label{Pott}
\end{equation}
Having the conserved and thermodynamic quantities of the rotating
dilaton black string at hand, we are in a position to check the
first law of black hole thermodynamics. Numerical calculations
show that the conserved quantities calculated above satisfy the
first law of thermodynamics
\begin{equation}
dM=TdS+\Omega d{J}+Ud{Q}.
\end{equation}

In conclusion, we constructed a new class of charged rotating
solutions in four-dimensional Einstein-Maxwell-dilaton gravity
with cylindrical or toroidal horizons in the presence of dilaton
potentials and investigate their properties. These solutions are
asymptotically anti-de Sitter. The cosmological constant couples
the dilaton field in a nontrivial way. The dilaton potential with
respect to the cosmological constant includes three Liouville-type
potentials. This is consistent with the arguments in \cite{MW}
that no (anti)-de-Sitter version of dilaton black holes exist with
only one Liouville-type dilaton potential. We found the suitable
counterterm which removes the divergences of the action in the
presence of three Liouville-type dilaton potential. We also
computed the conserved and thermodynamic quantities of the
solutions by using the conterterm method and verified numerically
that these quantities satisfy the first law of black hole
thermodynamics.

\acknowledgments{This work has been supported financially by
Research Institute for Astronomy and Astrophysics of Maragha,
Iran.}


\end{document}